\documentclass[12pt]{article}

\usepackage{graphicx}
\usepackage{adjustbox}

\begin{document}

\begin{center}

{\large\bf Exact enumeration approach to first-passage time distribution of non-Markov random walks}

Shant Baghram,$^1$, Farnik Nikakhtar,$^{1,2}$, M. Reza Rahimi Tabar,$^{1,3,\dagger}$ Sohrab Rahvar,$^{1,4}$ Ravi K.
Sheth,$^2$ Klaus Lehnertz,$^{5,6,7}$ and Muhammad Sahimi$^{8,\ddagger}$

{\it $^1$Department of Physics, Sharif University of Technology, Tehran 11155-9161, Iran\\ $^2$Center for
 Particle Cosmology, University of Pennsylvania, 209 S. 33rd Street, Philadelphia, PA 19104, USA\\
$^3$Institute of Physics, Carl von Ossietzky University of Oldenburg, Carl von Ossietzky Stra\ss{}e 9-11,
26111 Oldenburg, Germany\\
$^4$ Department of Physics, College of Science,
Sultan Qaboos University, P.O. Box 36, P.C. 123, Muscat, Sultanate of Oman\\
$^5$Department of Epileptology, University of Bonn, Sigmund Freud Stra\ss{}e 25, 53105 Bonn, Germany\\
$^6$Helmholtz Institute for Radiation and Nuclear Physics, University of Bonn, Nussallee 14-16, 53115
Bonn, Germany\\
$^7$Interdisciplinary Center for Complex Systems, University of Bonn, Br\"uhler Stra\ss{}e 7, 53175
Bonn, Germany\\
$^8$Mork Family Department of Chemical Engineering and Materials Science, University of Southern
California, Los Angeles, California 90089-1211, USA}

\end{center}

\bigskip

We propose an analytical approach to study non-Markov random walks by employing an exact enumeration method.
Using the method, we derive an exact expansion for the first-passage time (FPT) distribution for any
continuous, differentiable non-Markov random walk with Gaussian or non-Gaussian multivariate
distribution. As an example, we study the FPT distribution of a fractional Brownian motion with a Hurst
exponent $H\in(1/2,1)$ that describes numerous non-Markov stochastic phenomena in physics, biology and
geology, and for which the limit $H=1/2$ represents a Markov process.
\newpage

\begin{center}
{\bf I. INTRODUCTION}
\end{center}

The concept of first passage refers to the crossing of a prespecified location, or some sort of a
threshold, in a stochastic trajectory [1]. The distribution of the first-passage times (FPTs), which
represents the probability of crossing the trajectory at a specific time or location [2,3] and depends
on the nature of the stochastic process, plays a fundamental role in the theory of stochastic processes,
as well as in their applications. The FPT distribution makes it possible to investigate quantitatively
the uncertainty in the properties of a stochastic system within a finite time. Two important applications
are the extinction time of a disease in the models of epidemic phenomena, and the time for a species to
reach a critical threshold in population dynamics. In addition, the statistics of the FPT distribution
have many applications to diffusion-limited processes in physics [1], chemistry [4], and biology [5],
spreading of electrical blackouts [6], epidemiology [7], and even foraging animals [8,9], as well as
to understanding transport processes in disordered materials [10], porous media [11,12], neuroscience
[13-16], spreading of computer viruses [17], target search processes [18], economics [19], mathematical
finance [20,21], psychology [22], cosmology [23,24], and the reliability theory [25]. Through a suitable
boundary the FPT presents the first time that the error in the so-called clock model [26] becomes too
large and uncontrollable. Rapid detection of anomalies is closely related to recognizing the optimal
stopping time of a diffusion process [27] and, hence, the FPT distribution.

Due to their very large number of applications, the FPT properties have been studied extensively, and are
well understood when the stochastic phenomena represent a Markov process. As a general rule, however, the
dynamics of a given stochastic process in complex media is the result of its interactions with the
environment around it, which may contain trapping sites, obstacles, moving parts, active pumps, etc.
[28], and cannot be described as a Markov process. Indeed, although the evolution of the set of all
microscopic degrees of freedom of a system is Markovian, the dynamics restricted only to the random
walker is not [3,29,30]. Experimental realizations of non-Markov dynamics include diffusion of tracers in
crowded narrow channels [31] and in complex fluids, such as nematics [32] and viscoelastic solutions
[33,34], as well as the dark matter halo mass function [35]. Even in simple fluids, hydrodynamic memory
influences various phenomena and, thus, non-Markov dynamics has been reported recently [36].

Using inclusion-exclusion principle and an exact enumeration method, we derive in this paper the FPT
distribution of a non-Markov random walk by assuming that the trajectory of the walk is differentiable
at every point. As an example, we derive the FPT distributions of fractional Brownian motion (FBM) with
a given Hurst exponent $H\in(0.5,1)$. The analytical results are confirmed by extensive numerical
simulation and the analysis of $10^6$ trajectories.

The rest of this paper is organized as follows. In the next section we describe the exact enumeration
approach to derive the FPT probability density of a non-Markovian random walk. We then drive in Sec. III
an analytical expression for the FPT distribution of FBM. The results of numerical simulations are
presented in Sec. IV, while the paper is summarized in Sec. V. In the Appendix, we provide the details of
the derivation of our results.

\begin{center}
{\bf II. EXACT ENUMERATION METHOD FOR THE FPT PROBABILITY DENSITY}
\end{center}

We define a general dynamical equation for a random walk, $x(t)$, driven by a correlated, nonstationary
noise (velocity) $v(t)$,
\begin{equation}
\frac{\partial x(t)}{\partial t} = v(t)\;,\quad\quad C(t,t')=\langle v(t)~v(t')\rangle\;.
\end{equation}
$x(t)$ is assumed to be continuous and its derivative (velocity) $v(t)$ to be well-defined at any time
[37]. The noise $v(t)$ has a zero mean and an arbitrary $n$-point joint distribution $p(v_n,t_n;\cdots
v_1,t_1;v_0,t_0)$. The correlation function $C(t,t')$ depends on {\it both} $t$ and $t'$. Because $x(t)$
is a stochastic process, each of its realizations reaches a given barrier $x=x_c$ for the first time at
a different time $t$, giving rise to a FPT probability density $f(t)$. Consider the trajectories with the
initial conditions $x(t_0)=x_0$ and $\dot{x}(t_0)=v(t_0)=v_0$, crossing the barrier $x_c$ in the time
interval $t$ and $t+{\mathrm d}t$ with $v(t)>0$. The crossing is equivalent to the conditions that $x(t)<x_c$ and
$x_c<x(t+{\mathrm d}t)$ [16,38]. If $x_c$ is constant, $x(t)$ will lie in the interval $x_c-v{\mathrm d}t<x(t)<x_c$. Then,
the probability that $x(t)$ satisfies the passage condition $x_c-v{\mathrm d}t<x(t)<x_c$ is
\begin{displaymath}
\int_{x_c-vdt}^{x_c}P(x,v,t|x_0,v_0,t_0){\mathrm d}x=vP(x_c,v,t|x_0,v_0,t_0){\mathrm d}t\;,
\end{displaymath}
where we kept the terms up to the order of ${\mathrm d}t$. Since $v(t)>0$ at $x_c$, we should integrate over all
positive velocities. Therefore, the probability of crossing the barrier $x_c$ per unit time is given by
[16],
\begin{equation}
n_1(x_c,t_1|x_0,v_0,t_0)=\int_0^\infty v P(x_c,v,t_1|x_0,v_0,t_0){\mathrm d}v\;.
\end{equation}
Equation (2) represents the rate of up-crossing, rather than a density function and, thus, it is not
normalized. We generalize Eq. (2) to the joint probability of multiple up-crossings, i.e., $x(t)$
crossing the barrier in each of the intervals $(t_1,t_1+{\mathrm d}t),\cdots,(t_p,t_p+{\mathrm d}t)$, by integrating over all
the crossing points $t_1,t_2,\cdots,t_p$,
\begin{equation}
n_p(x_c,t_p;\cdots;x_c,t_1|x_0,v_0,t_0)=\int_0^\infty {\mathrm d}v_p\cdots\int_0^\infty {\mathrm d}v_1 v_p\cdots v_1
P(x_c,v_p,t_p;\cdots;x_c,v_1,t_1|x_0,v_0,t_0)\;.
\end{equation}
Using Bayes' theorem, one may substitute the conditional probability density in Eq. (3) with the joint
probability density. In Fig. 1 typical trajectories, as well as the FPT distribution of the FBM for
$x_c=1$ with $x_0=0$ are presented. The trajectories are constructed using the Cholesky decomposition
(see below).

A trajectory can cross $x_c$ several times (see the lower panel of Fig. 1). We relate the FPT
distribution to the statistical properties of the up-crossings, which are considered as point processes
with rates $n_p$, where $p$ refers to the number of up-crossing. To this end, we look for the fraction
of all the trajectories that up-cross $x_c$ for the first time at time $t$ with the initial conditions
$(x_0,v_0)$ at time $t_0$, and enumerate them in terms of $n_p$. To simplify the notation, we drop $x_c$
and the initial conditions.

The rate $n_1(t)$ is over-counted through the trajectories that had an up-crossing at shorter times
$t_1< t$. Therefore, we subtract their fraction from the first term. This stems from the fact that
$n_1(t)$ is a local function in $t$, but there is no guaranty that a trajectory has not up-crossed before
$t$. The over-counting implies that the main problem is a combinatorial counting. Thus, as an enumeration
technique we use the inclusion-exclusion principle, one of the most useful principles of counting in
combinatorics and probability. According to De Morgan's laws, in the general and complementary form, the
principle of inclusion-exclusion for finite sets $A_1,\;A_2,\dots,\;A_n$ is expressed by
\begin{equation}\label{comp-IEP}
\Big|\bigcap_{i=1}^n \bar{A}_i\Big|=\Big|U-\bigcup_{i=1}^n A_i\Big|=|U|-\sum_{i=1}^n|A_i|+\sum_{1\leq i
\leq j\leq n}|A_i\cap A_j|-\dots+\sum_{1\leq i\leq j\leq\cdots\leq n}(-1)^{n-1}|A_i\cap A_j\cap\dots\cap
A_n|\;,
\end{equation}
where $U$ is a finite universal set containing all the $A_i$, and $\bar{A_i}$ are the complement of $A_i$
in $U$. That the trajectories cross $x_c$ for the first time at time $t$ implies that they should not
have been crossed at $x_c$ at shorter times. We consider $n_1(t)$ as the universal set, and define the next
subset by $A_i=n_2(t,t_i)$, denoting the fraction of trajectories for which the up-crossing at time $t$
is not for the first time, and that they had a previous up-crossing at a shorter time $t_i<t$. Then, the
FPT distribution is given by, $|\bigcap_{i=1}^n\bar{A}_i|$, because only the trajectories that have a
first up-crossing at time $t$ and do not belong to the subsets $A_i$ are of interest. Using Eq. (4), we
obtain the FPT distribution [16]:
\begin{eqnarray}\label{eq5}
{f}(t) &=& \Big|\bigcap_{i=1}^n\bar{A}_i\Big|=n_1(t)-\int_0^t n_2(t,t_1){\mathrm d}t_1+\frac{1}{2!}\int_0^t
\int_0^t n_3(t, t_2, t_1){\mathrm d}t_1{\mathrm d}t_2 -\dots\cr \nonumber \\
&=& \displaystyle\sum_{p=0}^{\infty} \frac{(-1)^p}{p!}\int_0^t \dots \int_0^t n_{p+1}(t,t_p,\dots,t_1)
{\mathrm d}t_p \dots {\mathrm d}t_1\;,
\end{eqnarray}
where $n_{p+1}(t,t_p,\dots,t_1)$ are given by the conditional probabilities (3). The factor $1/p!$
accounts for the number of permutations of the variables $t_p,\dots,t_1$, with the signs explained in
Table I. To calculate $n_p(t_p,\dots, t_1)$, we consider the trajectories in the absence of $x_c$ and
let them return after an up-crossing and, then, up-cross the barrier $p$ times. The correct counting of
such multiple crossings yields the distribution $f(t)$ of the FPT. Equation (5) provides us with the
exact expansion of the FPT distribution for any continuous, differentiable non-Markov random walk with
Gaussian or non-Gaussian multivariate distribution [16]. We note that a naive truncation of the series would
give rise to a non-normalized (diverging in the long-time limit) distribution [39].

Let us define as a point process the time scales at which the trajectories cross $x_c$. The distributions
of such a point process are the aforementioned rate functions. Since the trajectories have nonzero
velocities, successive up-crossings cannot be too close, so that $n_p(t_p,\dots,t_1)$ is zero if two of
its arguments are equal. Such point processes represent systems of nonapproaching random points [40].
There are two types of decoupling approximations to deal with the infinite series in Eq. (5), which are
based on approximating the higher-order terms by the lower-order ones and are known as the Hertz and
Stratonovich approximations. The general expression for $f(t)$ is given by
\begin{equation}\label{fup-close form}
f(t) = \psi'(t) e^{-\psi(t)}\;.
\end{equation}
The Hertz approximation is based on assuming that all the up-crossings are independent of each other, and
that the correlations between them are negligible. This leads to the following FPT distribution with
$\psi_{\rm Hertz}(t)=\int_0^t n_1(t'){\mathrm d}t'$ [24,39,41]:
\begin{equation}\label{eq7}
f(t) \approx n_1(t) ~ \exp[-\int_0^t n_1(t'){\mathrm d}t']\;.
\end{equation}
In the {\it Hertz approximation} $n_p(t_p,\dots,t_1)$ factorizes to $n_1(t_p)\dots,n_1(t_1)$. In the
Stratonovich approximation, we calculate exactly the first and the second terms of the expansion and
approximate all the higher-order terms by the first two [35], with the corresponding FPT distribution
being in the form of Eq. (6) with [39],
\begin{equation}\label{eq9}
\psi_{\rm Str}(t)=-\int_0^tn_1(t')\frac{\ln{[1-\int_0^t R(t,t')n_1(t'){\mathrm d}t']}}{\int_0^tR(t,t')n_1(t'){\mathrm d}t'}
{\mathrm d}t'\;
\end{equation}
where $R(t_i,t_j)=1-n_2(t_i,t_j)/[n_1(t_i)n_1(t_j)]$. For simplicity and in order to derive an expression
for $f(t)$, we assume in the following that the velocity distribution is Gaussian.

\begin{center}
{\bf III. ANALYTICAL DERIVATION OF THE FPT DISTRIBUTION OF THE FBM}
\end{center}

We now derive the FPT distribution of the FBM with a Hurst exponent $H\in(0.5,1)$, which is defined in
terms of its nonstationary correlation function [42]:
\begin{equation}\label{fbmdef2}
\langle x_H(t_1)x_H(t_2)\rangle=\frac{1}{2}(|t_1|^{2H}+|t_2|^{2H}-|t_2-t_1|^{2H})\;,
\end{equation}
which is positive semidefinite (see the Appendix) with its first derivative (velocity) being the
fractional Gaussian noise (FGN) $v_H(t)$ so that, $\dot{x}_H(t)=v_H(t)$. Using physical arguments
[43,44], as well as rigorous analysis [45], it was shown that the scaling behavior of the FPT
distribution of a FBM has the following long-time behavior
\begin{equation}
f(t) \sim t^{H-2}\;.
\end{equation}
Given that the FBM and FGN have Gaussian distributions for $x$ and $v$, respectively, we determine
$n_1(t)$ and $n_2(t_1,t_2)$ and, therefore, $R(t_1,t_2)$ and the FPT distribution in the Hertz and
Stratonovich approximations. It is straightforward to show that $n_1(t)$ is given by the following
expression,
\begin{equation}
n_1(t)=p(x_c)\int_0^\infty v p(v|x_c){\mathrm d}v\;,
\end{equation}
where $p(v|x_c)$ is a Gaussian distribution with mean $\langle v|x_c\rangle=x_c\langle vx\rangle/
\langle x^2\rangle=x_cH/t$ and variance $s_{v|x_c}=H^2t^{2H-2}/\Gamma^2$, where $\Gamma^2=
\gamma^2/(1-\gamma^2)$ and $\gamma^2=\langle xv\rangle^2/\langle x^2\rangle\langle v^2\rangle$. For the
FBM, $\langle xv\rangle^2=H^2t^{4H-2}$ and $\langle x^2\rangle=t^{2H}$. In the Appendix, we present an
expression for $\langle v^2\rangle$ in terms of the Hurst exponent $H$. We find that the explicit
expression for $n_1(t)$ is given by
\begin{displaymath}
n_1(t)=\frac{\Gamma^2}{2\pi Ht^{2H-1}}\exp{\left(-\frac{y^2}{2}\right)}\left\{\frac{H^2 t^{2H-2}}
{\Gamma^2}\exp{\left(-\frac{y^2\Gamma^2}{2}\right)}+\frac{H^2x_c}{2\Gamma}t^{H-2}\sqrt{2\pi}
\left[1+{\rm erf}\left(\frac{y\Gamma}{\sqrt{2}}\right)\right]\right\}\;,
\end{displaymath}
where $y=x_c / t^H$. Using Eq. (7) we obtain the FPT distribution in the Hertz approximation, which, in general, is
accurate for estimating the first peak of the FPT distribution, but it over- or underestimates its tail.
Similarly, we find that,
\begin{eqnarray}
n_2(t_1,t_2) & = & \int_0^\infty v{\mathrm d}v \int_0^\infty {\mathrm d}v' v' p(x_c,x_c',v,v')\cr \nonumber \\
& = & p(x_c) \int_0^\infty {\mathrm d}v v p(v,x_c) p(x_c'|x_c,v)\int_0^\infty {\mathrm d}v' v' p(v'|x_c', x_c,v)\;,
\label{n2}
\end{eqnarray}
where all the distributions in Eq. (12) are Gaussian. For example, $p(x_c'|x_c,v)$ has the mean (see the
Appendix for the variance)
\begin{eqnarray*}
\langle x_c'|x_c,v\rangle=x_c\frac{\langle x'x\rangle}{t^{2H}}+(v-\langle v|x_c\rangle)\frac{\langle
x'v\rangle-\langle x'x\rangle/2t^{2H}}{s_{v|x_c}}\;.
\label{meanx}
\end{eqnarray*}
The correlation functions $\langle x'x\rangle$ and $\langle x'v\rangle$ are given by Eq. (A.9) in the
Appendix, and $s_{v|x_c}=H^2t^{2H-2}/\Gamma^2$. Having $n_1(t)$ and $n_2(t_1,t_2)$ enables one to
determine $R(t_1,t_2)$ and $f(t)$ in the Hertz and Stratonovich approximations.

\begin{center}
{\bf IV. NUMERICAL RESULTS}
\end{center}

Figure 1 presents the trajectories of a FBM process using the Cholesky decomposition [46,47] (see the Appendix) and their
FPT distribution for $H=0.8$, $x_c=1$, and $x_0=0$. In Fig. 2 the FPT distributions of the FBM
trajectories is plotted. The FPT is obtained from the Cholesky method.
%for $x_c=1$ with $x_0=0$ are presented, which were calculated using the trajectories that
%were computed via the Cholesky decomposition.
In these plots, we also show the FPT distributions in the Hertz approximation, which deviate from the FPT directly computed using trajectories. For comparison, the theoretically-predicted tails of the distributions, i.e., $f(t)\sim t^{H-2}$, are also plotted. Figure 2
indicates that the theoretical tails of the FPT in the long-time limit coincide with the FPT distributions computed
using the trajectories. As already mentioned above the Hertz approximation predicts correctly the location of the peak
of the FPT distribution, but underestimates the tails.

To derive the FTP distribution in the Stratonovich approximation with $H =0.8$, one must calculate
$\psi_{\rm Str}(t)$ via Eq. (9), and then use Eq. (6). To avoid any error from the numerical
differentiation of $\psi_{\rm Str}(t)$, we determine the integrated FPT distribution via the term
$\exp[-\psi_{\rm Str}(t)]$. In Fig. 3 the cumulative FPT distribution is presented for $H=0.6$ and
$H=0.8$, indicating that the Hertz approximation deviates clearly from the results computed via the
Cholesky decomposition. As shown in Fig. 2, the tail of $f(t)$ in the Hertz approximation does not
coincide completely with those obtained by the Cholesky decomposition. Higher-order approximations, e.g.,
the Stratonovich approximation, are therefore needed, implying that $n_2(t_i,t_j)$ should not be
factorized as $n_1(t_1)n_1(t_2)$.  As shown in Fig. 3, the Stratonovich approximation provides better
estimations for the FPT distributions.

One may define various measures to study the interdependence of the up-crossing events. The simplest
measure is the Fano factor. Consider a time window $T$ and count the mean number (and its variance) of
up-crossing events for trajectories in the window. The Fano factor ${\cal F}(T)$ is defined as the
variance of the number of up-crossing events in $T$, divided by its mean number, and is written in terms
of $n_2(t_1,t_2)$ and $n_1(t)$ [48]. More specifically, the Fano factor is given by ${\cal F}=\langle
\Delta N^2\rangle/\langle N\rangle$ (with $\langle\Delta N^2\rangle=\langle N^2\rangle-\langle
N\rangle^2$), where $\langle N\rangle=\int_0^T n_1(t){\mathrm d}t$ and $\langle N^2\rangle=\langle N\rangle+
\int_0^T\int_0^T n_2(t_2,t_1){\mathrm d}t_2{\mathrm d}t_1$ [48]. For independent point processes, i.e., $n_2(t_2,t_1)=
n_1(t_2) n_1(t_1)$, one has ${\cal F}= 1$. Therefore, for a Poisson process ${\cal F}(T)=1$. By
definition, ${\cal F}(T)>1$ and ${\cal F}(T)<1$ refer, respectively, to over- and under-dispersion [49].
We plot in Fig. 4 the Fano factor versus the size of the time window $T$, which indicates that, in the
long-time limit, the up-crossing point processes are strongly over-dispersed. This means that in such
time scales $n_2(t_i,t_j)$ should not be factorized as $n_1(t_1)n_1(t_2)$, and that the Hertz
approximation is not appropriate for estimating the tails of the FPT distribution. A very crucial point to indicate is
that in the time span which ${\cal F} \sim 1$ the Hertz approximation works well and it is very near to the Cholesky-derived  FPT distribution. On the other and if
the Fano factor deviates from unity, it is certain that the Hertz approximation is not suitable for FPT, however this parameter can not quantify the accuracy of  Stratonovich approximation.

\begin{center}
{\bf V. SUMMARY}
\end{center}

Except for the limiting case of Markov processes, no exact analytical expression for the FPT distribution
of general non-Markov random walks had been derived. In principle, the FPT distribution of non-Markov
processes may be obtained from the solution of the associated Fokker-Planck equation with absorbing
boundaries in higher dimensions, resulting from the Markovian embedding of a non-Markov process [50].
Even the calculation of the mean FPT for a non-Markov process is, however, a rather difficult task, since
the corresponding boundary problem cannot be treated in a straightforward manner [51-55]. We presented
a general method for deriving such analytical expressions %, and employed it to derive one (as a series)
for the FPT distribution. This is done by using an exact enumeration method based on combinatorics and the
inclusion-exclusion principle, which can be generalized to include the FTP distribution of non-Markov
random walks in higher dimensions. As an example, analytical results were presented for the FBM with the
Hurst exponent $H\in(0.5,1)$, which is a non-Markov process with infinite-range memory, and has
wide applications in many disciplines [28]. The numerical results were also compared with two well-known
approximations, namely, the Hertz and Stratonovich approximations, which revealed their shortcomings.

\begin{center}
{\bf ACKNOWLEDGMENTS}
\end{center}
S.B. acknowledges the partial support of Sharif University of Technology, Grant No. G960202 for this paper. F.N. acknowledges support from the National Science Foundation Graduate Research Fellowship under Grant No. DGE-1845298.

\bigskip
\begin{center}
{\bf APPENDIX}
\end{center}

\setcounter{equation}{0}
\renewcommand{\theequation}{A.{\arabic{equation}}}

\bigskip

We provide the details of the main results presented in the main text of the paper.

\begin{center}
{\bf A. Variance of the velocity of fractional Brownian motion}
\end{center}

The time derivative (increments) of the FBM is the FGN, and has the following correlation function
\begin{equation}
C_H (\tau,\delta) = \frac{\sigma^2\delta^{2H-2}}{2}\left[\left(\frac{|\tau|}{\delta}+1\right)^{2H}+
\left|\frac{|\tau|}{\delta} -1\right|^{2H}-2\left|\frac{\tau}{\delta}\right|^{2H}\right]\;,
\end{equation}
where $0<H<1$, and $\tau=t_2-t_1$. Here, $\delta>0$ is used for smoothing the FBM to make it numerically
differentiable [42]. We note that in the limit $\tau\to 0$, the $\delta$-dependence of $\gamma^2=\langle
xv\rangle^2/\langle x^2\rangle\langle v^2\rangle$ drops out. In the literature [42], there is no unique
expression for $\langle v^2\rangle$. Here, by generating the FBM trajectories and numerically
differentiating them for $H\in(0.5,1)$, the best fit is found to be $\langle v^2\rangle=c_0+c_1 H^m$,
where $c_0=-2.47\pm 0.01$, $c_1=2.88\pm 0.05$, and $ m=-4.72\pm 0.02$.

\begin{center}
{\bf B. Fractional Gaussian noise}
\end{center}

The stochastic representation of the FBM is given by,
\begin{equation}
B_H (t)=B_H (0) + \frac{1}{\Gamma(H+1/2)}\left\{\int_{-\infty}^0\left[(t-s)^{H-1/2}-(-s)^{H-1/2}\right]
\,{\mathrm d}W(s)+\int_0^t (t-s)^{H-1/2}\,{\mathrm d}W(s)\right\}\;,
\end{equation}
where $dW(s)$ is a Wiener process that is written in terms of the Gaussian white noise $\xi(s)$ as,
$dW(s)=\xi(s)ds$. The FGN is then defined by, $G_H(t)=dB_H(t)/dt$. Taking the time derivative of Eq.
(A2) yields
\begin{eqnarray}
G_H(t) &=& \frac{1}{\Gamma(H+1/2)}\left\{\int_{-\infty}^0(H-\frac{1}{2})(t-s)^{H-3/2}{\mathrm d}W(s)+\left[
(t-s)^{H-1/2}\xi(s)\frac{{\mathrm d}}{{\mathrm d}t}t\right]\bigg|_{s=t}\right\}
\cr \nonumber \\ &+&  \frac{1}{\Gamma(H+1/2)}\left[\int_0^t(H-\frac{1}{2})(t-s)^{H-3/2}{\mathrm d}W(s)\right]\;.
\end{eqnarray}
The second term on the r.h.s. of  Eq. (A3) is not finite for $H\in(0,0.5)$. Therefore, the FBM has
no well-defined "velocity" for the Hurst exponent in the range of $(0,0.5)$.

\begin{center}
{\bf C. Proof for the variance of the FBM being positive semidefinite}
\end{center}

A symmetric $n\times n$ real matrix {\bf C} is the covariance of some random (Gaussian) vector, if and
only if it is positive semidefinite, which means that
\begin{equation}
{\bf z}'{\bf Cz}=\sum_{i=1}^n\sum_{i=j}^n z_iz_jC_{i,j}\geq 0\;\;\;\forall\;z_1,\cdots z_n\in{\rm I\!R}
\;,
\end{equation}
where $z$ here is the aforementioned random vector. The FBM has a vanishing mean ($x(0)=0$), while its covariance is given by Eq. (10) of the main text, for
$(t_1,t_2) \geq 0$ and $H\in(0,1)$. We show that
\begin{equation}
C(t_1,t_2) = \frac{1}{2}(|t_1|^{2H}+|t_2|^{2H}-|t_2-t_1|^{2H})\;,
\end{equation}
is a covariance function. Consider the function
\begin{equation}
\Phi(t_2, r) = (t_2-r)^{\alpha_+ - 1/2} - (-r)^{\alpha_+ - 1/2}\;,
\end{equation}
defined for all $t_2\geq 0$ and $r\in{\rm I\!R}$, where $\alpha_+ = {\rm max}(0,H)$ for all $H\in{\rm
I\!R}$. Since $H <1$, we can determine $\int_{-\infty}^\infty|\Phi(t_2, r)|^2{\mathrm d}r <\infty$ and
\begin{equation}
\int_{-\infty}^{\infty}\Phi(t_2,r)\Phi(t_1,r){\mathrm d}r=\kappa C(t_1,t_2)\;\;\; \forall ~(t_1, t_2) \geq 0\;,
\end{equation}
where $\kappa$ is a positive and finite constant that depends only on $H$. Therefore, we find
\begin{eqnarray}
\sum_{i=1}^n \sum_{j=1}^n z_i z_j C_{t_i,t_j} &=& \frac{1}{\kappa}\sum_{i=1}^n \sum_{j=1}^n z_i z_j
\int_{-\infty}^\infty\Phi(t_i,r)\Phi(t_j,r){\mathrm d}r \cr \nonumber &=& \frac{1}{\kappa}\int_{-\infty}^{\infty}
\left[\sum_{j=1}^n z_i \Phi(t_i,r)\right]^2{\mathrm d}r \geq 0\;.
\end{eqnarray}

\begin{center}
{\bf D.  Analytical expressions for ${\bf n_1(t)}$ and ${\bf n_2(t,t')}$ with a Gaussian velocity}
\end{center}

Due to the linearity of the system, all the joint probability densities are Gaussian and have the form
\begin{equation}
P_{n}({\bf Q})=\frac{1}{(2\pi)^{n/2}\sqrt{\det\hat{{\bf C}}_{n}}}\exp\left(-
\frac{{\bf Q}\hat{{\bf C}}_{n}^{-1}{\bf Q}}{2}\right).
\label{E:GaussProbab}
\end{equation}
Here, ${\bf Q}=(q_{1}(t_{1}),\dots,q_{n}(t_{n}))$ is an $n$-dimensional vector whose $i$th component is
the coordinate $x(t_i)$ or the velocity $v(t_i)$ at time $t_i$, and $\hat{{\bf C}}_{n}$ is the symmetric
$n\times n$ correlation matrix whose entries are the correlation functions between the corresponding
components of the vector ${\bf Q}$: $C_{ij}=C_{ji}=\langle q_{i}(t_{i})q_{j}(t_{j})\rangle$. Then,
$n_1(t)$ is obtained in closed analytical form:
\begin{equation}
%\begin{split}
n_1(t)=\frac{\Gamma^2}{2\pi Ht^{2H-1}}\exp\left(-\frac{y^2}{2}\right)\left\{\frac{H^2 t^{2H-2}}
{\Gamma^2}\exp\left(-\frac{y^2\Gamma^2}{2}\right)+\frac{H^2 x_c}{2\Gamma}t^{H-2}\sqrt{2\pi}
\left[1+\mathrm{erf}\left(\frac{y\Gamma}{\sqrt{2}}\right)\right]\right\}\;,
%\end{split}
\label{E:1termExpl}
\end{equation}
where $y= x_c / t^H$ and $\Gamma^2=\gamma^2/(1-\gamma^2)$. For the joint densities of multiple up-crossings $n_p(t_p,\dots,
t_1)$ no closed expression can be obtained. We evaluate the integral over $v_1$ in Eq. (3) analytically
and then perform numerical integration of the resulting expression over $v_2,\dots,v_p$ to determine
$n_p(t_p,\dots,t_1)$. The integrals over time in the expressions for $f(t)$ are also evaluated
numerically. For $n_2(t, t')$, we compute the mean and variance of the conditional distributions,
\begin{equation}
%\begin{split}
n_2(t,t')=\int_0^\infty v{\mathrm d}v\int_0^\infty {\mathrm d}v' v'p(x_c,x'_c,v,v')=p(x_c)\int_0^\infty {\mathrm d}vv p(v|x_c)
p(x'_c|x_c,v)\int_0^\infty {\mathrm d}v' v' p(v'|x'_c,x_c,v)\;.
%\end{split}
\end{equation}
Assuming that $t'> t$, the correlations are given by
\begin{eqnarray}
%\begin{split}
\langle x'x\rangle & = & \frac{1}{2}\left[t'^{2H}+t^{2H}-(t'-t)^{2H}\right]\\
\langle v'x\rangle & = & Ht'^{2H-1}-H(t'-t)^{2H-1} \\
\langle x'v\rangle & = & Ht^{2H-1}+H(t'-t)^{2H-1} \\
\langle v'v\rangle & = & H(2H-1)(t'-t)^{2H-2}\;,
%\end{split}
\end{eqnarray}
where, for example, $\langle x'x\rangle=\langle x(t')x(t)\rangle$. We also know that
\begin{equation}
\sigma^2_x = \langle x^2 \rangle = t^{2H}, \hspace{30pt} \langle x v \rangle = Ht^{2H-1}\;.
\end{equation}
Note that all the conditional distributions are Gaussians and, therefore, they are specified by their
mean and variance. For example, for $p(x'_c|x_c,v)$ we have
\begin{eqnarray}
%\begin{split}
\mu_{x'|x, v} & = & \langle x'|x,v\rangle=\langle x'|x\rangle+\frac{\langle(x'-\langle x'|x\rangle)(v-
\langle v|x\rangle)\rangle}{\sigma^2_{v|x}}(v-\langle v|x\rangle) \nonumber\\
& = & \frac{t'^{2H} + t^{2H} - (t' - t)^{2H}}{2t^{2H}}x_c + \frac{\Gamma^2}{H^2 t^{2H-2}}\nonumber\\
& \times & \left\{Ht^{2H-1}+H(t'-t)^{2H-1}-\frac{H}{2t}\left[t'^{2H}+t^{2H}-(t'-t)^{2H}\right]\right\}
(v-\frac{H}{t}x_c )\;,
%\end{split}
\end{eqnarray}
\begin{eqnarray}
%\begin{split}
\sigma^2_{x'|x,v} & = & \sigma^2_{x'}-\frac{\langle xx'\rangle^2}{\sigma^2_x}-\frac{1}{\sigma^2_{v|x}}
\left[\langle x'v\rangle-\frac{\langle x'x\rangle\langle xv\rangle}{\sigma^2_x}\right]^2 \nonumber \\
& = & t'^{2H}-\frac{[t'^{2H} + t^{2H} - (t'-t)^{2H}]^2}{4t^{2H}}\nonumber \\
& - & \frac{\Gamma^2}{H^2t^{2H-2}}\left\{Ht^{2H-1}+H(t'-t)^{2H-1}-\frac{H}{2t}\left[t'^{2H}+t^{2H}
-(t'-t)^{2H}\right]\right\}^2\;.
%\end{split}
\end{eqnarray}
For $p(v'|x',x,v)$, we should calculate the mean and variance of $p(v'|x, v)$, which are given by
\begin{eqnarray}
%\begin{split}
\langle v'|x,v\rangle & = & \frac{Ht'^{2H-1}-H(t'-t)^{2H-1}}{t^{2H}}x_c+\frac{\Gamma^2}{H^2t^{2H-2}}
\nonumber\\
& \times & \left\{H(2H-1)(t'-t)^{2H-2}-\frac{H}{t}\left[Ht'^{2H-1}-H(t'-t)^{2H-1}\right]\right\}
(v-\frac{H}{t}x_c)\;,
%\end{split}
\end{eqnarray}
\begin{eqnarray}
%\begin{split}
\sigma^2_{v'|x,v} & = & \sigma^2_{v'} (1 - \frac{\langle x v' \rangle^2}{\sigma^2_{x}\sigma^2_{v'}}) -
\frac{1}{\sigma^2_{v|x}}(\langle v v'\rangle-\frac{\langle v'x\rangle\langle xv\rangle}{\sigma^2_x})^2
\nonumber\\
& = & \sigma^2_{v'}\left(1-\frac{\langle xv'\rangle^2}{\sigma^2_{x}\sigma^2_{v'}}\right)\nonumber\\
& - & \frac{\Gamma^2}{H^2t^{2H-2}}\left\{H(2H-1)(t'-t)^{2H-2}-\frac{H}{t}\left[Ht'^{2H-1}-H(t'-t)^{2H-1}
\right]\right\}^2\;.
%\end{split}
\end{eqnarray}
Now, for $p(v'|x', x, v)$ we obtain
\begin{eqnarray}
%\begin{split}
\mu_{v'|x',x,v} & = & \langle v'|x',x,v\rangle=\langle v'|x,v\rangle+\frac{1}{\sigma^2_{x'|x,v}}\Bigg[
\langle x'v'\rangle \nonumber\\
& - & \frac{1}{1 - \gamma^2}(\frac{\langle v' x \rangle \langle x x'\rangle}{\sigma^2_x} +
\frac{\langle v'v\rangle \langle v x' \rangle}{\sigma^2_v})\\
& - & \frac{\langle v' x \rangle \langle x v \rangle \langle v x' \rangle}{\sigma^2_x \sigma^2_v}
- \frac{\langle v' v \rangle \langle v x \rangle \langle x x'\rangle}{\sigma^2_x \sigma^2_v})\Bigg]
(x_c-\langle x'|x,v\rangle)\nonumber\;.
%\end{split}
\end{eqnarray}
\begin{eqnarray}
%\begin{split}
\sigma^2_{v'|x',x,v} & = & \Bigg(1-\frac{1}{\sigma^2_{x'|x,v}\sigma^2_{v'|x,v}}\Bigg[\langle x'v'\rangle
\nonumber\\
& - & \frac{1}{1-\gamma^2} (\frac{\langle v' x \rangle \langle x x'\rangle}{\sigma^2_x} + \frac{\langle
v'v\rangle \langle v x' \rangle}{\sigma^2_v} \frac{\langle v' x \rangle \langle x v \rangle \langle v
x'\rangle}{\sigma^2_x \sigma^2_v} \\
& - & \frac{\langle v' v \rangle \langle v x \rangle \langle x x'\rangle}{\sigma^2_x \sigma^2_v} \Bigg]^2
\Bigg) \sigma^2_{v'|x, v}\nonumber\;.
%\end{split}
\end{eqnarray}

\begin{center}
{\bf E. The Cholesky decomposition}
\end{center}

To compute the non-Markovian first up-crossing distribution for the FBM, we must generate trajectories
with the correct ensemble properties. Here, we describe an algorithm to generate such trajectories.
Equation (9) of the main text defined, $C_{ij}\equiv C(t_i,t_j)=\langle x(t_i)x(t_j)\rangle$, the
correlation between the $x(t)$ between times $t_i$ and $t_j$. The matrix {\bf C} is real, symmetric, and
positive-definite and, therefore, it has a unique decomposition, ${\bf C}={\bf LL}^{\rm T}$, in which
{\bf L} is a lower triangular matrix, which is known as the Cholesky's decomposition. We use {\bf L} to
generate the ensemble of the trajectories as follows.

First, consider a vector $\mathbf{\xi}$, which is Gaussian white noise with zero mean and unit variance
(i.e. $\langle\xi_m\xi_n\rangle=\delta_{mn}$). If we generate the desired trajectories as
\begin{equation}
x(t_i) = x_i = \sum_j L_{ij}\, \xi_j,
\label{deltaL}
\end{equation}
then $x_i$ will have the correlations of random walk given by
\begin{equation}
\langle x_i x_j \rangle=\sum_{m,n}L_{im}L_{jn}\langle\xi_m\xi_n\rangle={\bf LL}^{\rm T}={\bf C}\;.
\end{equation}
Since {\bf L} is triangular, % each $x_i$ requires
only a sum over $j\le i$ is needed in matrix calculations, so the method is fast.
Next, we provide a proof of the Cholesky decomposition, and present it in terms of the correlation
matrix. In a more general context, there is a sufficient condition for a square matrix to have a LU
decomposition, ${\bf C}={\bf LU}$, where {\bf L} and {\bf U} are, respectively, the lower and upper
triangular matrices of {\bf C}. If all the $n$ leading principal minors of the $n\times n$ matrix {\bf C}
are nonsingular, then {\bf C} has an LU decomposition. Let us recall that the $k$th leading principle
minor of {\bf C} is given by
\begin{equation}
{\bf C}_k=\left(\begin{array}{cccc}
%\begin{pmatrix}
c_{11} & c_{12} & \cdots & c_{1k} \\
c_{12} & c_{22} & \cdots & c_{2k} \\
\vdots & \vdots & \vdots & \vdots \\
c_{1k} & c_{2k} & \cdots & c_{kk}
%\end{pmatrix}\;,
\end{array}\right)\;,
\end{equation}
where we have assumed that ${\bf C}_1,{\bf C}_2,\cdots,{\bf C}_n$ are nonsingular. Using induction, it
is not difficult to show that there is a LU decomposition for the correlation matrix. Using the symmetry
of {\bf C}, we write
\begin{equation}
{\bf LU}={\bf C}={\bf C}^{\rm T}={\bf U}^{\rm T}{\bf L}^{\rm T}\;,
\end{equation}
which implies that
\begin{equation}
{\bf U}({\bf L}^{\rm T})^{-1}={\bf L}^{-1}{\bf U}^{\rm T}\;.
\end{equation}
The l.h.s of the equation is upper triangular, whereas the r.h.s. is a lower triangular matrix.
Consequently, there is a diagonal matrix {\bf D} such that ${\bf D}={\bf U}({\bf L}^{\rm T})^{-1}$.
Then, ${\bf U}={\bf DL}^{\rm T}$, which for the correlation matrix implies that, ${\bf C}={\bf
LDL}^{\rm T}$, where {\bf D} is a positive-definite matrix with its elements also being positive.
Accordingly, we write {\bf C} as ${\bf C}=\tilde{{\bf L}}\tilde{{\bf L}}^{\rm T}$, where $\tilde{{\bf L}}
={\bf LD}^{1/2}$, which is the Cholesky decomposition.

It is clear that the matrix $\tilde{{\bf L}}$ is a lower triangular matrix as well, and can be used to
transform independent normal variables into dependent multinormal variables, which is the main idea of
the method we propose to construct the exact trajectories. The matrix $\tilde{{\bf L}}$ is calculated by
[40,41]
\begin{equation}
\tilde{{\bf L}}=\left(\begin{array}{ccccc}
%\begin{pmatrix}
1 & 0 & 0 & \cdots & 0 \\
c_{12} & \sqrt{1 - c_{12}^2} & 0 & \cdots & 0 \\
c_{13} & \frac{c_{23}-c_{12}c_{13}}{\sqrt{1-c_{12}^2}} & \sqrt{1 - c_3 R_2^{-1} c_3^T} & \cdots & 0 \\
\vdots & \vdots & \vdots & \vdots & \vdots \\
c_{1n} & \frac{c_{2n}-c_{12}c_{1n}}{\sqrt{1-c_{12}^2}} & \frac{c_{3n} - c_3^{*n} R_2^{-1} c_3^T }
{\sqrt{1 - c_3 R_2^{-1} c_3^T}} & \cdots & \sqrt{1 - c_n R_{n-1}^{-1}c_n^T}
\end{array}\right)\;,
%\end{pmatrix}\;,
\end{equation}
where, $R_n=c_{ij}|_{i,j=1}^n$ is a positive-definite correlation matrix, $R^{-1}$ is its inverse, and
$c_i^{*j}=(c_{1j},c_{2j},\dots,c_{i-1 j})$ for $j\geq i$, so that $c_i\equiv c_i^{*i}$. We note that for
a semi-positive definite matrix we should remove the first row and first column of the matrix in order
to have a positive-definite matrix, and then apply the Cholesky decomposition.

Algorithmically, our Cholesky decomposition algorithm constructs {\bf L} as follows:\\

{\tt input} $n, C_{ij}$\\

{\tt for} $k = 1, 2, ..., n$ {\tt do}

$\qquad L_{kk} \leftarrow \Big( C_{kk} - \displaystyle\sum_{s=1}^{k-1} L_{ks}^2\Big)^{1/2}$

$\qquad${\tt for} $i = k+1, k+2, ..., n$ {\tt do}

$\qquad\qquad L_{ik}\leftarrow\Big(C_{ik}-\displaystyle\sum_{s=1}^{k-1} L_{is}L_{ks}\Big)\Big/ L_{kk}$

$\qquad${\tt end}

{\tt end}\\

{\tt output} $L_{ij}$\\

\noindent All the trajectories for FBM in this paper were constructed using this algorithm.

\bigskip

\bigskip

$^\dagger$tabar@uni-oldenburg.de

$^\ddagger$moe@usc.edu

\newcounter{bean}
\begin{list}%
{[\arabic{bean}]}{\usecounter{bean}\setlength{\rightmargin}{\leftmargin}}

\item S. Redner, {\it A Guide to First-Passage Processes} (Cambridge University Press, Cambridge, 2001).

\item S. Condamin, O. B\'enichou, V. Tejedor, R. Voituriez, and J. Klafter, First-passage times in
complex scale-invariant media, Nature {\bf 450}, 77 (2007).

\item T. Gu\'erin, N. Levernier, O. B\'enichou, and R. Voituriez, Mean first-passage times of
non-Markovian random walkers in confinement, Nature {\bf 534}, 356 (2016).

\item S.A. Rice, {\it Diffusion-Limited Reactions}, vol. 25 (Elsevier, Amsterdam, 1985).

\item M.J. Saxton, A biological interpretation of transient anomalous subdiffusion. II. Reaction
kinetics, Biophys. J. {\bf 94}, 760 (2008).

\item B.A. Carreras, V.E. Lynch, I. Dobson, and D.E. Newman, Critical points and transitions in an
electric power transmission model for cascading failure blackouts, Chaos {\bf 12}, 985 (2002).

\item A.L. Lloyd and R.M. May, Epidemiology - how viruses spread among computers and people, Science
{\bf 292}, 1316 (2001).

\item O. Benichou, C. Loverdo, M. Moreau, and R. Voituriez, Two-dimensional intermittent search
processes: An alternative to L\'evy flight strategies, Phys. Rev. E {\bf 74}, 020102 (2006).

\item G.M. Viswanathan, E.P. Raposo, and M.G.E. da Luz, L\'evy flights and superdiffusion in the context
of biological encounters and random searches, Phys. Life Rev. {\bf 5}, 133 (2008).

\item D. Ben-Avraham and S. Havlin, {\it Diffusion and Reactions in Fractals and Disordered Systems}
(Cambridge University Press, London, 2000).

\item M. Sahimi, H.T. Davis, and L.E. Scriven, Dispersion in disordered porous media, Chem. Eng. Commun.
{\bf 23}, 329 (1983).

\item M. Sahimi, B.D. Hughes, L.E. Scriven, and H.T. Davis, Dispersion in flow through porous media: I.
One-phase flow, Chem. Eng. Sci. {\bf 41}, 2103 (1986).

\item H.C. Tuckwell, {\it Introduction to Theoretical Neurobiology} (Cambridge University Press, London,
1988).

\item A.N. Burkitt, A review of the integrate-and-fire neuron model: I. Homogeneous synaptic input, Biol.
Cybern. {\bf 95}, 1 (2006).

\item L. Sacerdote and M.T. Giraudo, in, {\it Stochastic Biomathematical Models}, edited by M. Bachar,
J. Batzel, and S. Ditlevsen (Springer, New York, 2013), p. 99.

\item T. Verechtchaguina, I.M. Sokolov, and L. Schimansky-Geier, First passage time densities in
non-Markovian models with subthreshold oscillations, Europhys. Lett. {\bf 73}, 691 (2006).

\item R. Pastor-Satorras and A. Vespignani, Epidemic spreading in scale-free networks, Phys. Rev. Lett.
{\bf 86}, 3200 (2001).

\item O. Benichou, M. Coppey, M. Moreau, P.-H. Suet, and R. Voituriez, Optimal search strategies for
hidden targets, Phys. Rev. Lett. {\bf 94}, 198101 (2005).

\item Q. Hu, Y. Wang, and X. Yang, The hitting time density for a reflected Brownian motion, Comput.
Econ. {\bf 40}, 1 (2012).

\item J. Janssen, O. Manca, and R. Manca, {\it Applied Diffusion Processes, from Engineering to Finance}
(Wiley, New York, 2013).

\item V. Linetsky, Lookback options and diffusion hitting times: A spectral expansion approach, Finance
Stoch. {\bf 8}, 373398 (2004).

\item D.J. Navarro and I.G. Fuss, Fast and accurate calculations for first-passage times in Wiener
diffusion models, J. Math. Psych. {\bf 53}, 222 (2009).

\item M. Musso and R.K. Sheth, The importance of stepping up in the excursion set approach, Mon. Notices
R. Astro. Soc. {\bf 438}, 2683 (2014).

\item M. Musso and R.K. Sheth, The excursion set approach in non-Gaussian random fields, Mon. Notices
R. Astr. Soc. {\bf 439}, 3051 (2014).

\item V. Pieper, M. Domin, and P. Kurth, Level crossing problems and drift reliability, Math. Methods
Oper. Res. {\bf 45}, 347 (1997).

\item C. Zucca and P. Tavella, The clock model and its relationship with the Allan and related variances,
IEEE Trans. Ultras. Ferroelectrics and Frequency Control {\bf 52}, 289 (2005).

\item C. Zucca, P. Tavella, and G. Peskir, Detecting atomic clock frequency trends using an optimal
stopping method, Metrologia {\bf 53}, S89 (2016).

\item I.M. Sokolov, Models of anomalous diffusion in crowded environments, Soft Matter {\bf 8}, 9043
(2012).

\item R. Friedrich, J. Peinke, M. Sahimi, and M.R. Rahimi Tabar, Approaching complexity by stochastic
methods: from biological systems to turbulence, Phys. Rep. {\bf 506} 87 (2011).

\item M. Anvari, M.R. Rahimi Tabar, J. Peinke, and K. Lehnertz, Disentangling the stochastic behavior of
complex time series, Sci. Rep. {\bf 6}, 35435 (2016).

\item Q.-H. Wei, C. Bechinger, and P. Leiderer, Single-file diffusion of colloids in one-dimensional
channels, Science {\bf 287}, 625 (2000).

\item T. Turiv, I. Lazo, A. Brodin1, B.I. Lev, V. Reiffenrath, V.G. Nazarenko, and O.D. Lavrentovich,
Effect of collective molecular reorientations on Brownian motion of colloids in nematic liquid crystals,
Science {\bf 342}, 1351 (2013).

\item D. Ernst, M. Hellmann, J. K\"ohler, and M. Weiss, Fractional Brownian motion in crowded fluids,
Soft Matter {\bf 8}, 4886 (2012).

\item T.G. Mason and D.A. Weitz, Optical measurements of frequency-dependent linear viscoelastic moduli
of complex fluids, Phys. Rev. Lett. {\bf 74}, 1250 (1995).

\item F. Nikakhtar, M. Ayromlou, S. Baghram, S. Rahvar, M.R. Rahimi Tabar, and R.K. Sheth, The excursion
set approach: Stratonovich approximation and Cholesky decomposition, Mon. Notices R. Astro. Soc.
{\bf 478}, 4, 5296 (2018).

\item T. Franosch, M. Grimm, M. Belushkin, F.M. Mor, G. Foffi, L. Forr\'o, and S. Jeney, Resonances
arising from hydrodynamic memory in Brownian motion, Nature {\bf 478}, 85 (2011).

\item M. Scott, {\it Applied Stochastic Processes in Science and Engineering} (Springer, Berlin, 2013).

\item G.R. Jafari, M.S. Movahed, S.M. Fazeli, M.R. Rahimi Tabar, and S.F. Masoudi, Level crossing
analysis of the stock markets, J. Stat. Mech. {\bf 06}, P06008 (2006).

\item T. Verechtchaguina, I.M. Sokolov, and L. Schimansky-Geier, First passage time densities in
resonate-and-fire models, Phys. Rev. E {\bf 73}, 031108 (2006).

\item R. Stratonovich, {\it Topics in the Theory of Random Noise}, Vol. 2 (Taylor \& Francis, London,
1967).

\item P. Hertz, Uber den gegenseitigen durchschnittlichen Abstand von Punkten, die mit bekannter mittlerer Dichte im Raume angeordnet sind, Mathematische Annalen {\bf 67}, 387 (1909).

\item B.B. Mandelbrot and J.W. van Ness, Fractional Brownian motions, fractional Gaussian noises and
applications, SIAM Rev. {\bf 10}, 422 (1968).

\item M. Ding and W. Yang, Distribution of the first return time in fractional Brownian motion and its
application to the study of on-off intermittency, Phys. Rev. E {\bf 5}, 207 (1995).

\item J. Krug, H. Kallabis, S.N. Majumdar, S. Cornell, A.J. Bray, and C. Sire, Persistence exponents for
fluctuating interfaces, Phys. Rev. E {\bf 56}, 2702 (1997).

\item G. M. Molchan, Maximum of a fractional Brownian motion: probabilities of small values, Commun.
Math. Phys. {\bf 205}, 97 (1999).

\item J.E. Gentle, {\it Numerical Linear Algebra for Applications in Statistics} (Springer, Berlin,
1998), p. 93.

\item V. Madar, Direct formulation to Cholesky decomposition of a general nonsingular correlation matrix,
Stat. Probab. Lett. {\bf 103}, 142 (2015).

\item T. Engel, {\it Firing Statistics in Neurons as Non-Markovian First Passage Time Problem}, Ph.D.
Thesis, Humboldt-Universit\"at zu Berlin (2006).

\item D.R. Cox and V. Isham, {\it Point Processes} (Chapman \& Hall, London, 1980).

\item P. H\"anggi, P. Talkner, and M. Borkovec, Reaction-rate theory: fifty years after Kramers, Rev.
Mod. Phys. {\bf 62}, 251 (1990).

\item C.R. Doering, P.S. Hagan, and C.D. Levermore, Bistability driven by weakly colored Gaussian
noise: The Fokker-Planck boundary layer and mean first-passage times, Phys. Rev. Lett. {\bf 59}, 2129
(1987).

\item P. H\"anggi, P. Jung, and P. Talkner, Comment on ``Bistability driven by weakly colored Gaussian
noise: The Fokker-Planck boundary layer and mean first-passage times,'' Phys. Rev. Lett. {\bf 60}, 2804
(1988).

\item R. Graham and T. T\'el, Nonequilibrium potential for coexisting attractors, Phys. Rev. A {\bf 33},
1322 (1986).

\item M.I. Dykman, P.V.E. McClintock, V.N. Smelyanski, N.D. Stein, and N.G. Stocks, Optimal paths and
the prehistory problem for large fluctuations in noise-driven systems, Phys. Rev. Lett. {\bf 68}, 2718
(1992).

\item S.M. Soskin, Large fluctuations in multiattractor systems and the generalized Kramers problem,
J. Stat. Phys. {\bf 97}, 609 (1999).

\end{list}%

%\end{document}

\newpage

\newpage

\bigskip
{Table I: The exact enumeration method. The $n$th column corresponds to the $n$th term of the sum in Eq.
(5).}
\begin{center}

\scalebox{0.9}{
\begin{tabular}{ c | c | c | c | c}
 & $|U|$ & $- \displaystyle\sum_{i=1}^n |A_i|$ & $ +\displaystyle\sum_{1 \leq i \leq j
\leq n} |A_i \cap A_j|$ & $ - \displaystyle\sum_{1 \leq i \leq j \leq k \leq n} |A_i \cap A_j \cap A_k|$
\\[20pt]
\hline \hline
& $= n_1(t)$ & $- \int_0^t n_2(t_i, t) {\mathrm d}t_i$ & $+ \frac{1}{2!}\int_0^t\int_0^t n_3(t_i, t_j, t) {\mathrm d}t_i
{\mathrm d}t_j$ & $- \frac{1}{3!}\int_0^t\int_0^t\int_0^t n_4(t_i, t_j, t_k, t) {\mathrm d}t_i {\mathrm d}t_j {\mathrm d}t_k$ \\
& \includegraphics[width=0.25\textwidth]{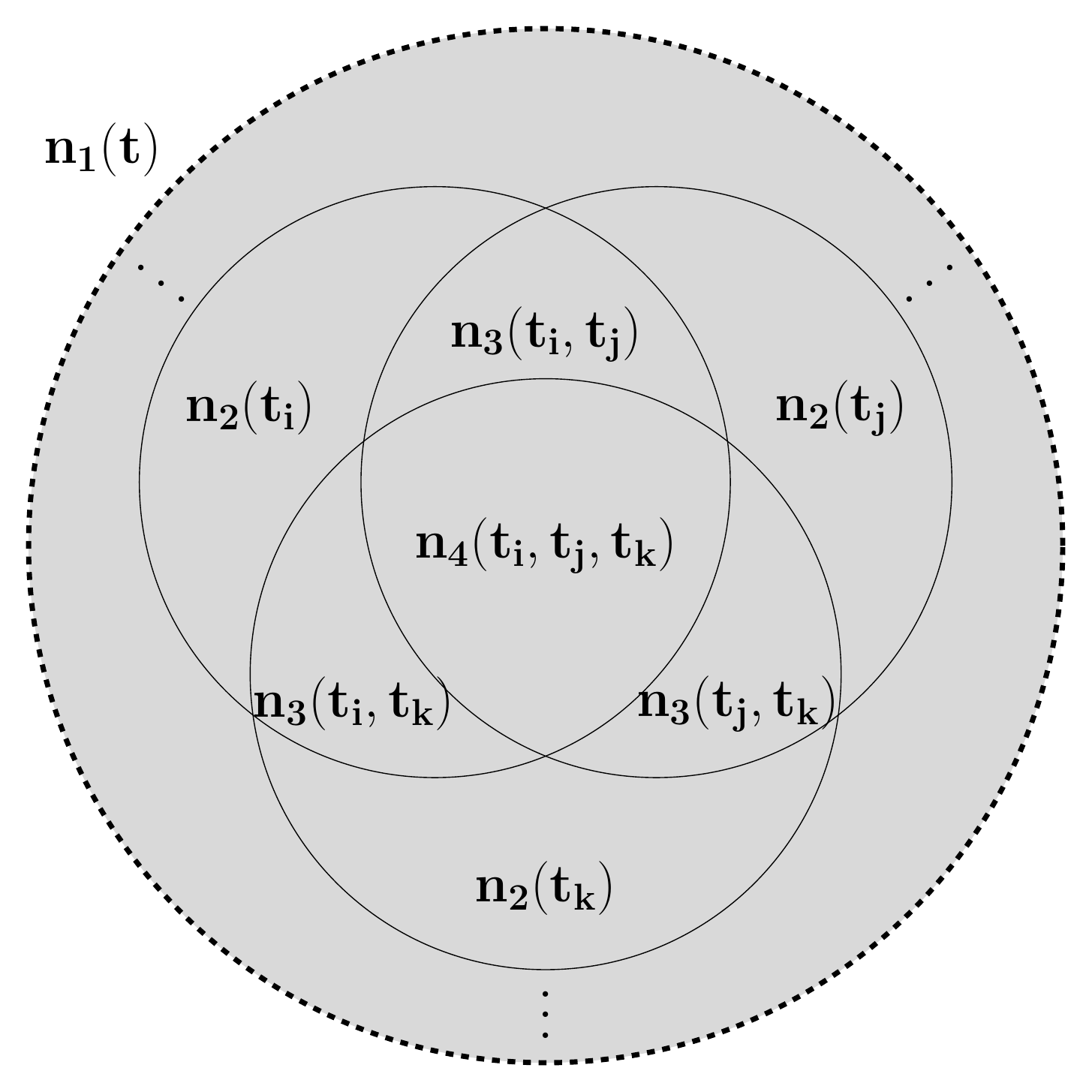} & \includegraphics[width=0.25\textwidth]{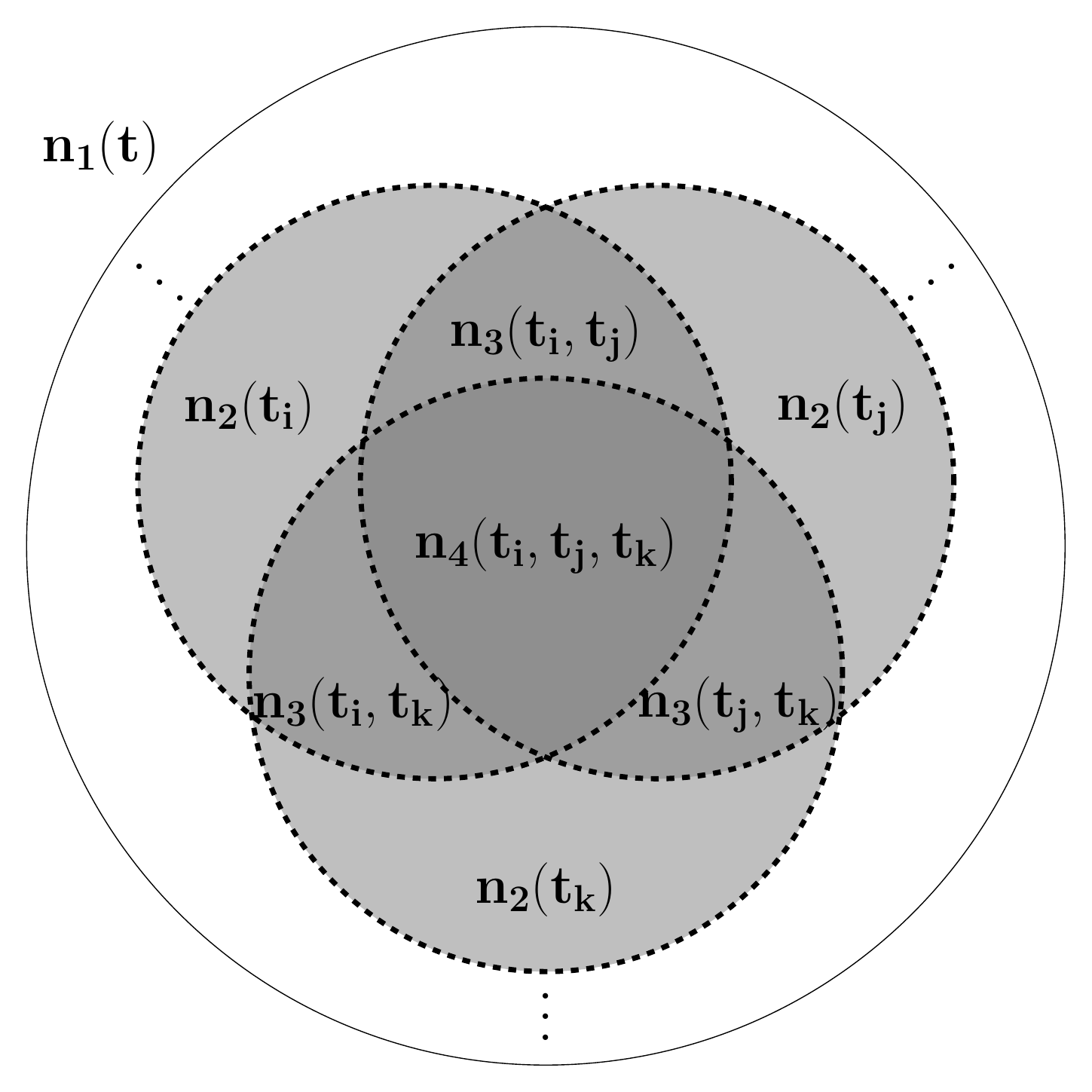} &
\includegraphics[width=0.25\textwidth]{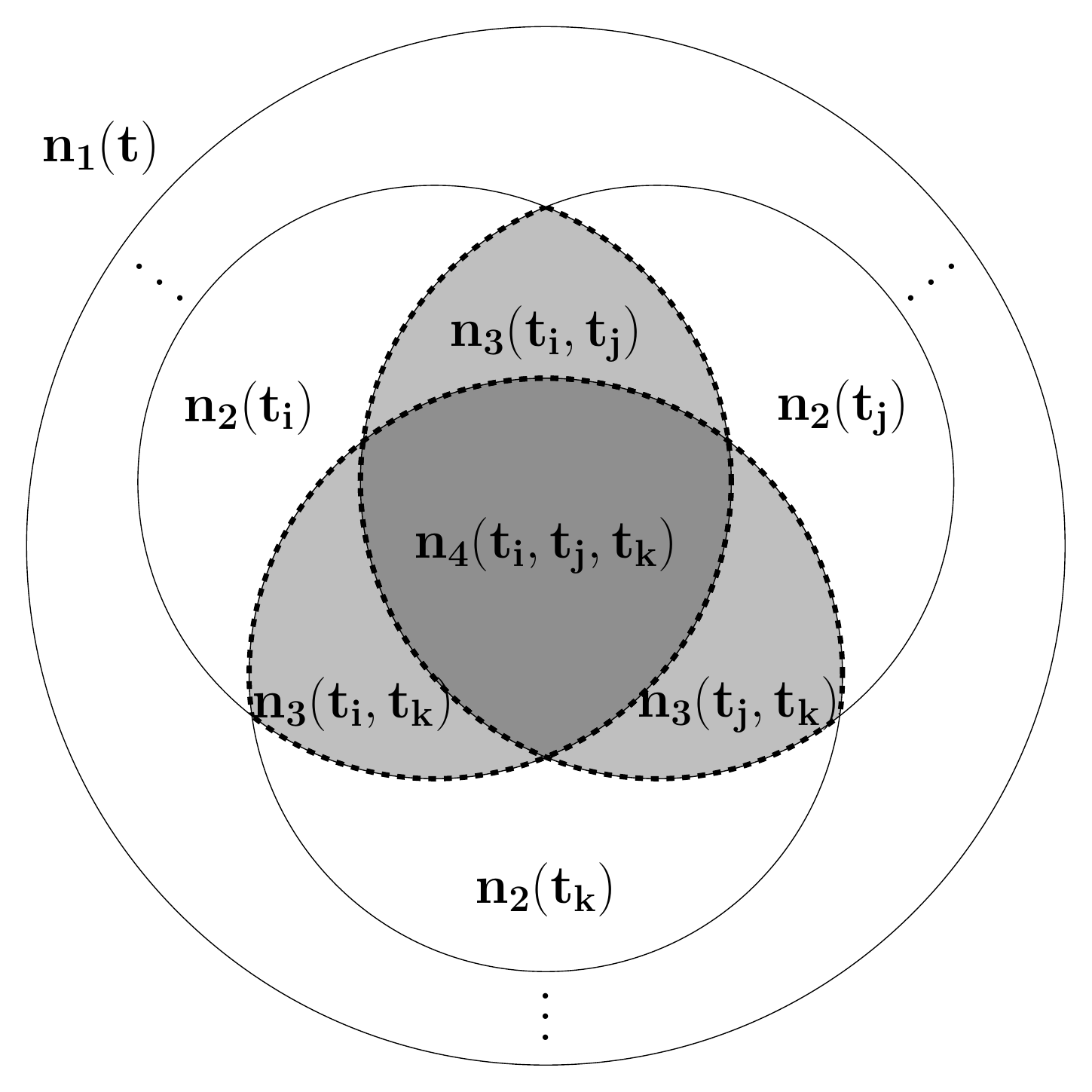} & \includegraphics[width=0.25\textwidth]{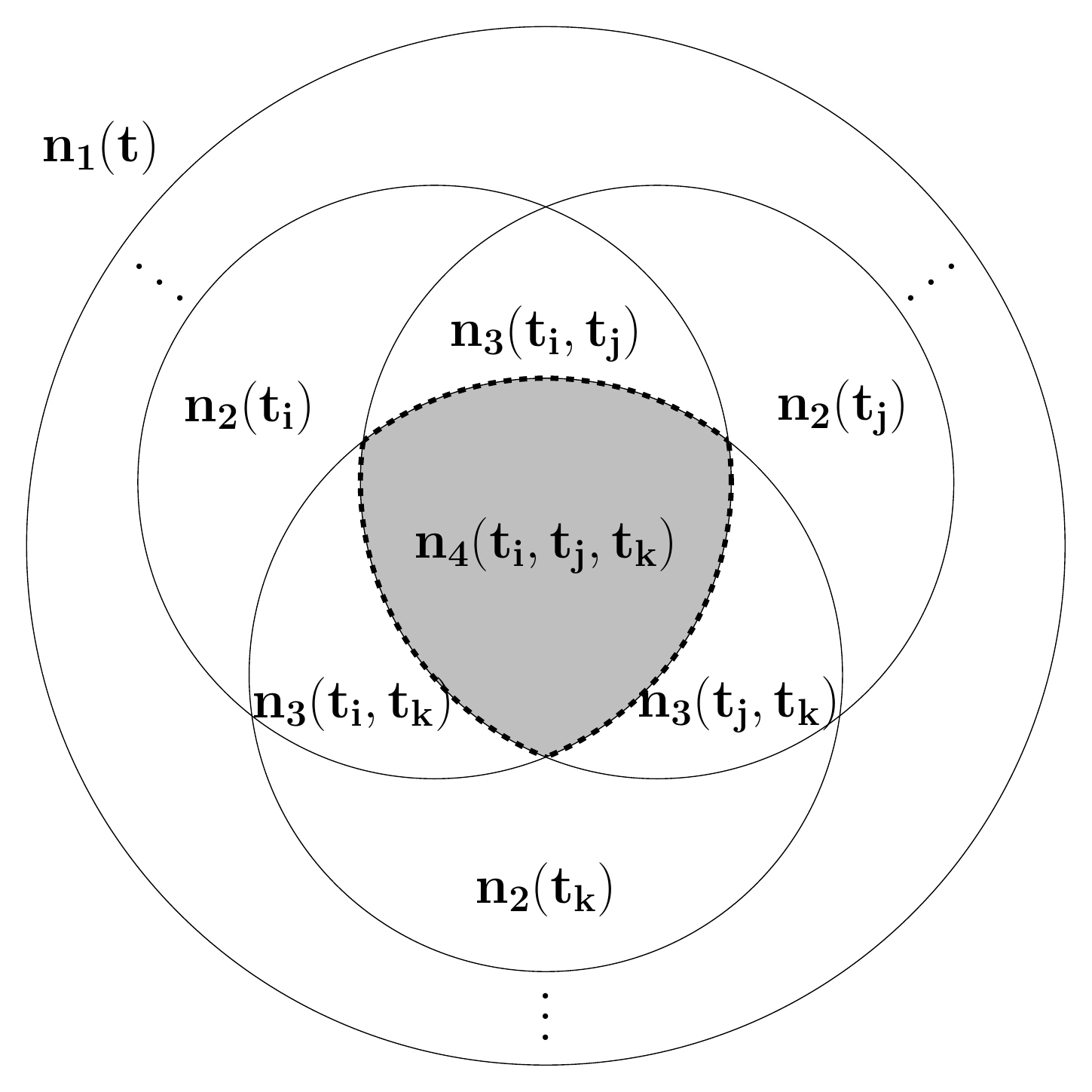} \\
\end{tabular}
}
\end{center}

\begin{center}
\begin{figure}
\includegraphics[width=0.6\columnwidth]{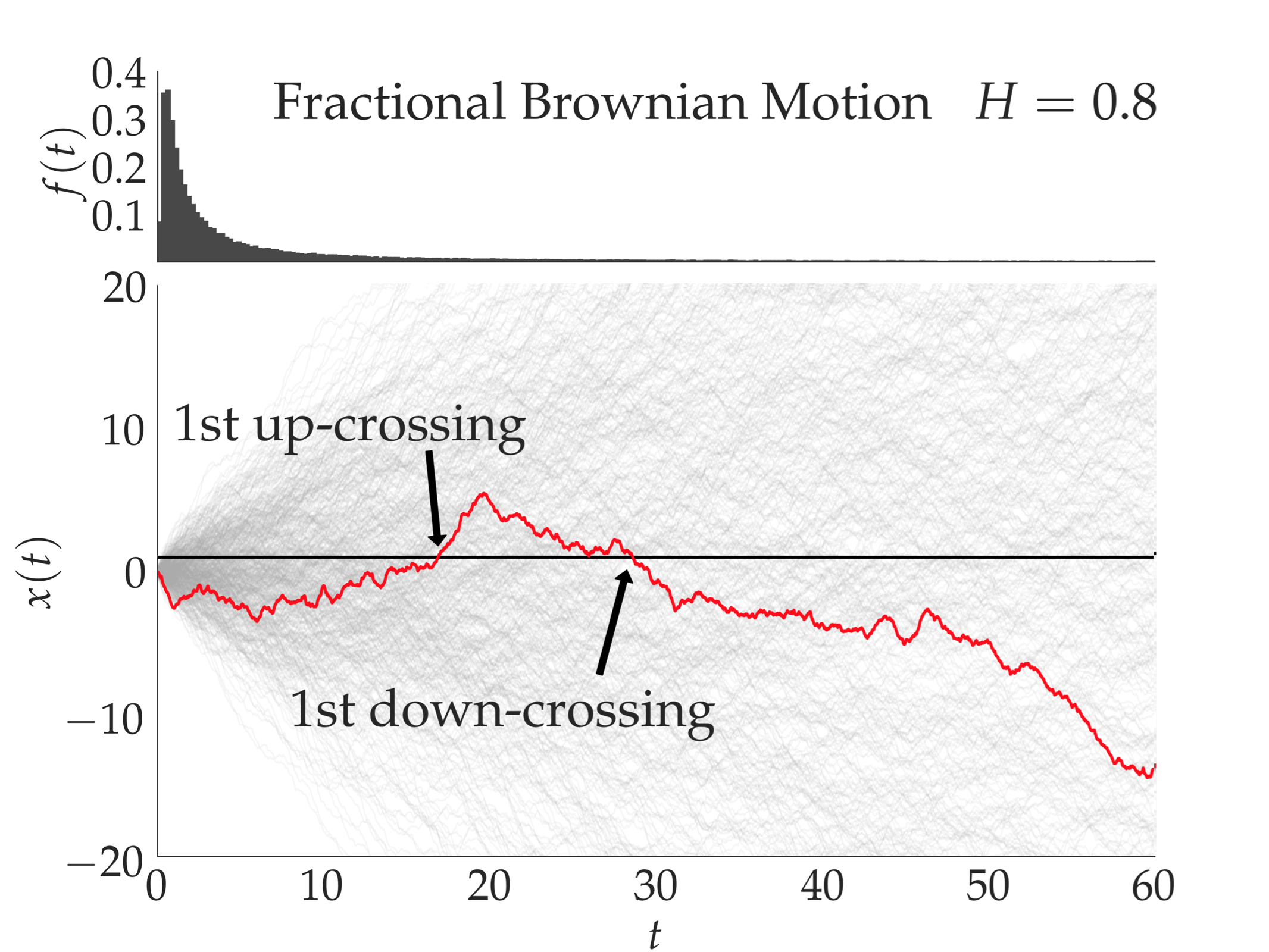}
\caption{\footnotesize{Sample trajectories of a non-Markov random walk with its corresponding FPT
distribution. Shown are the trajectories, as well as the FPT distribution, of the fractional Brownian
motion with the barrier $x_c=1$ with $x_0=0$. Trajectories were computed via the Cholesky decomposition.
The first crossings are marked with arrows for one trajectory.}}
\label{fig:fig1}
\end{figure}
\end{center}

\newpage

\begin{figure}
\includegraphics[width=0.49\columnwidth]{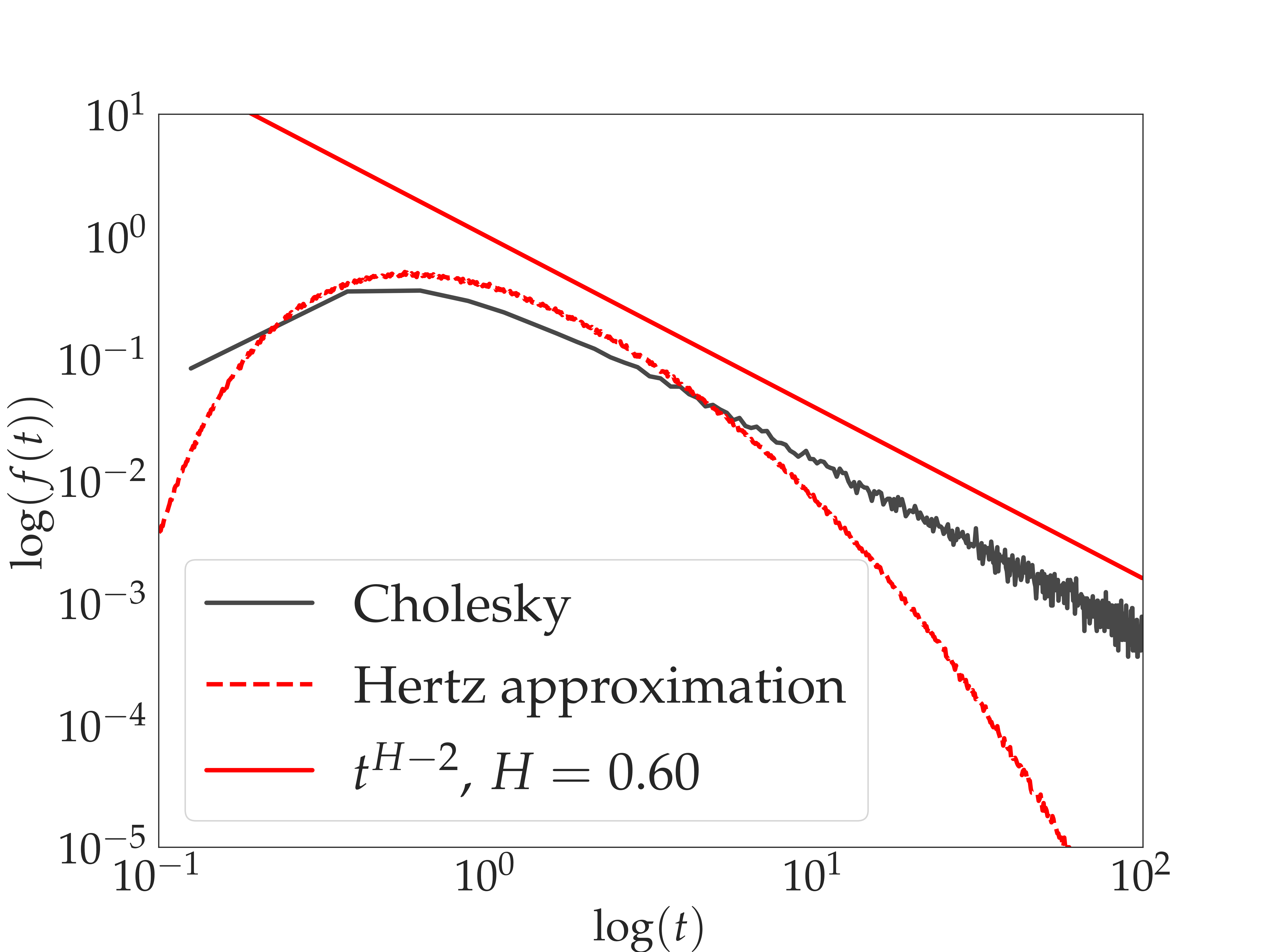}
\includegraphics[width=0.49\columnwidth]{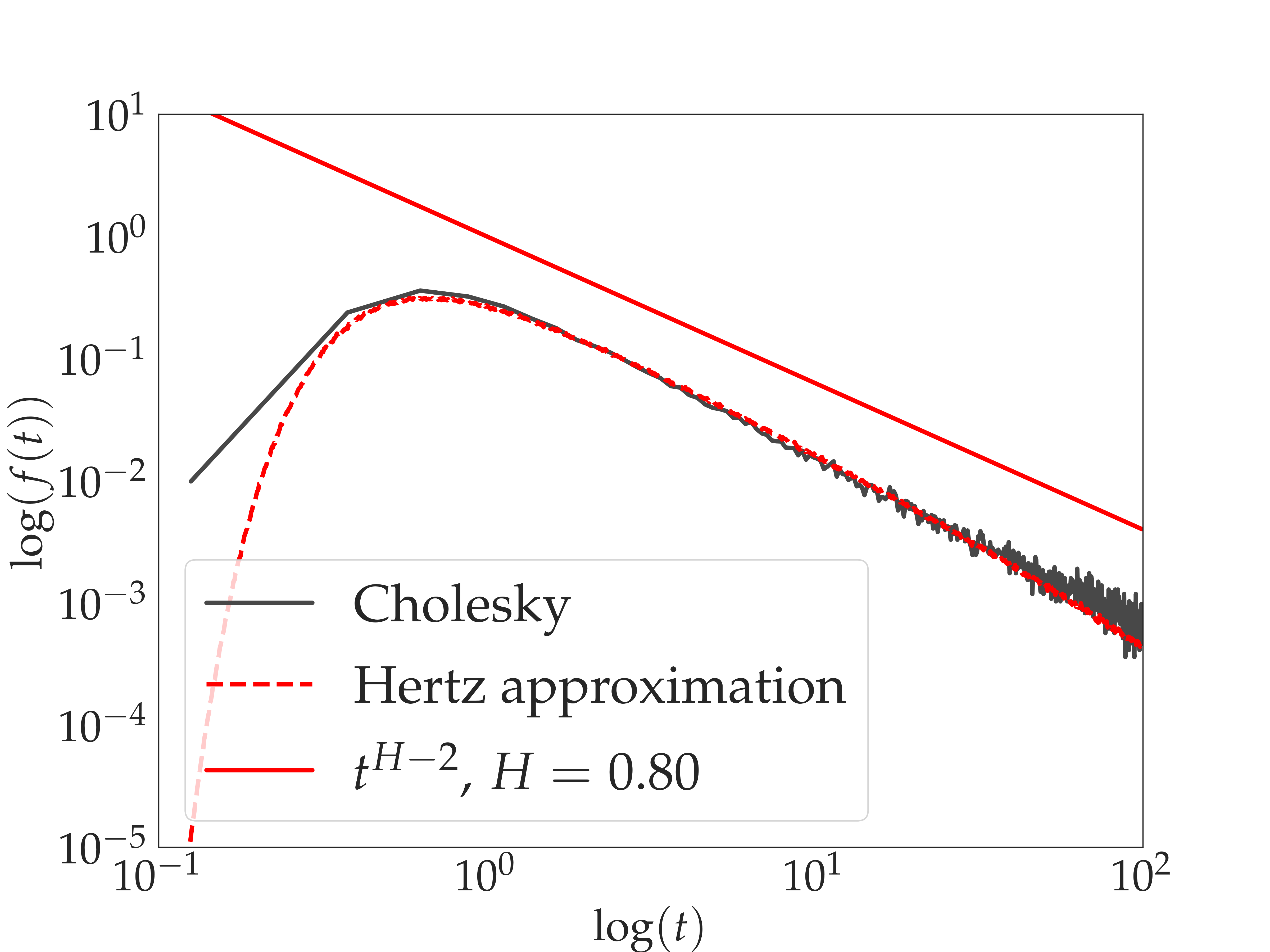}
\caption{\footnotesize{Comparison of the FPT distribution computed by using the trajectories and the
Cholesky decomposition, with the one obtained by the Hertz approximation, for the barrier $x_c=1$ with
$x_0=0$. For comparison, the theoretically-predicted tail of the distribution, i.e., $f(t)\sim t^{H-2}$,
is also shown [45].}}
\label{fig:fig2}
\end{figure}

\newpage

\begin{center}
\begin{figure}
\includegraphics[width=0.49\columnwidth]{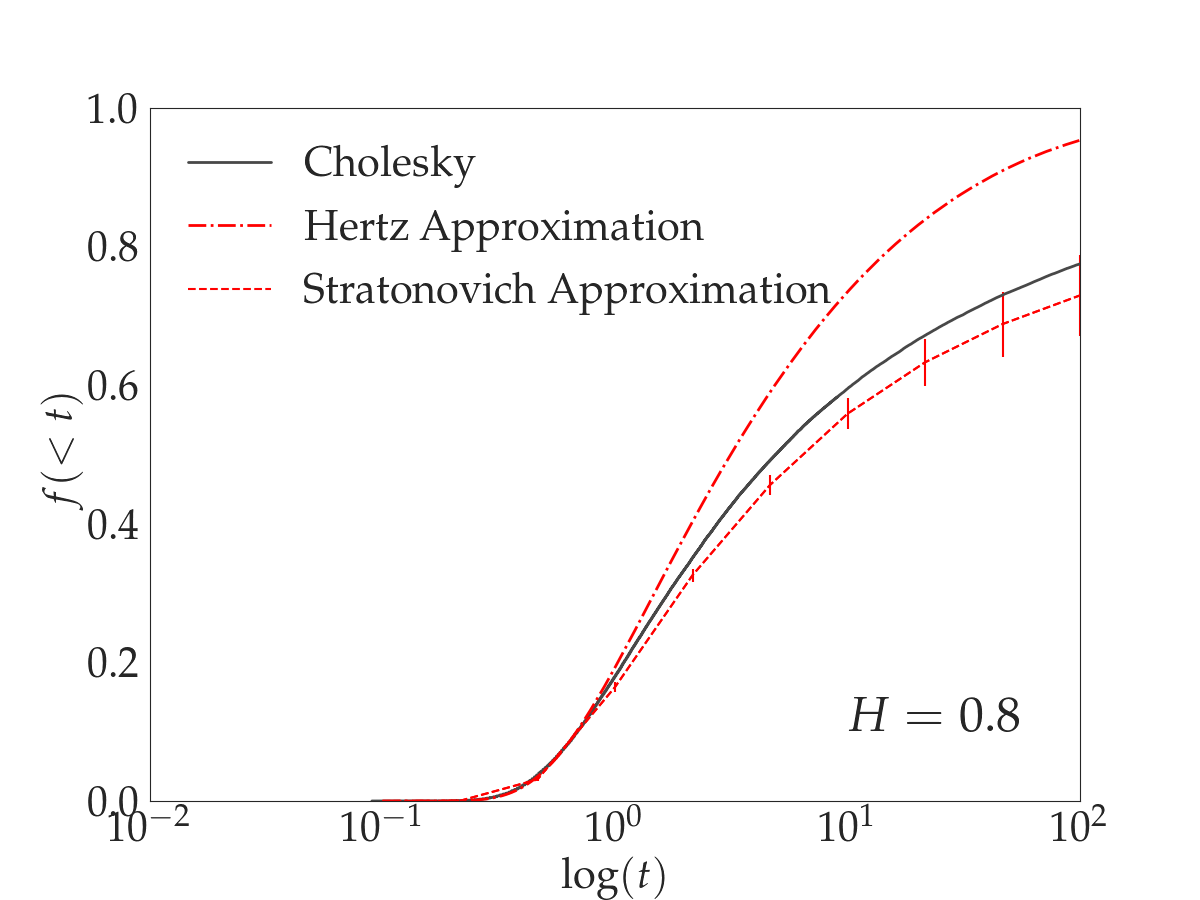}
\includegraphics[width=0.49\columnwidth]{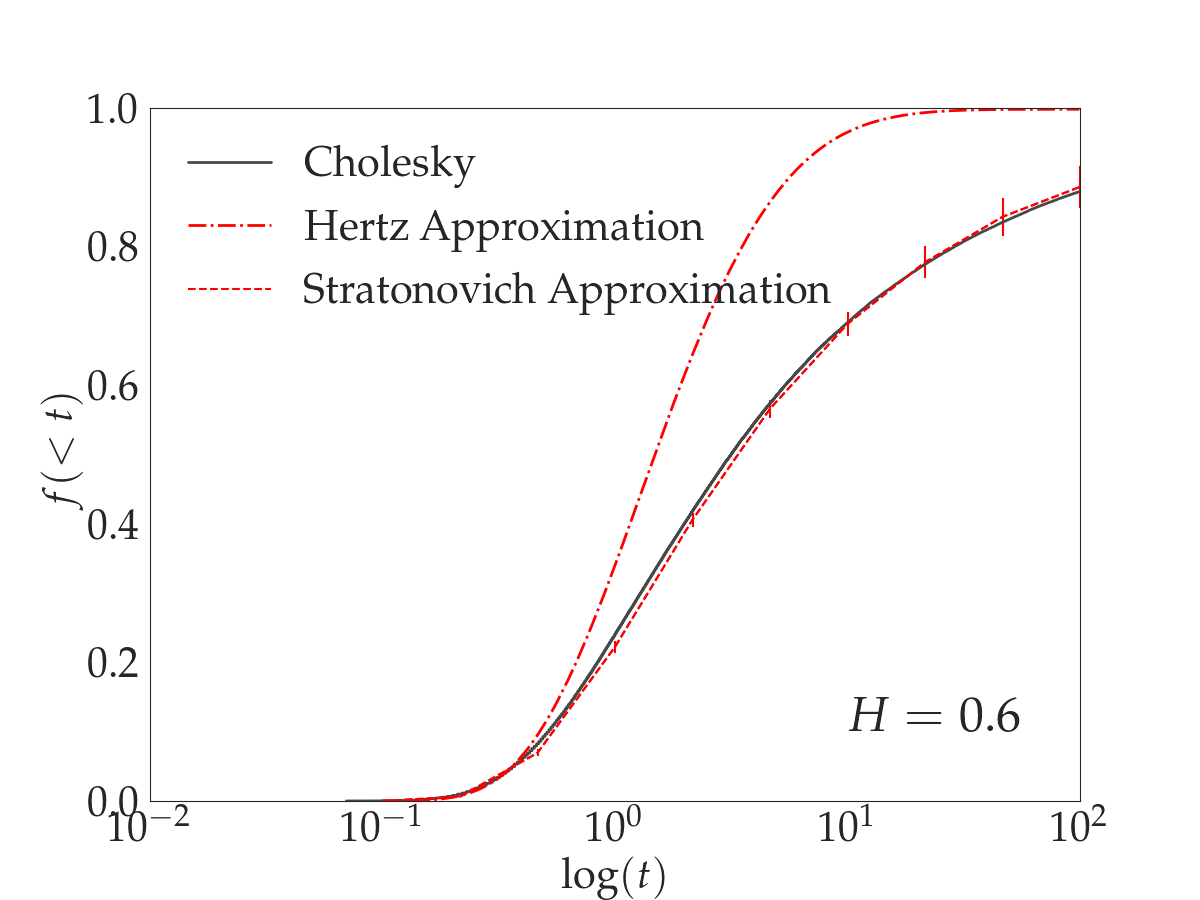}
\caption{\footnotesize{The cumulative FPT distributions in the Hertz (red dot-dashed lines) and
Stratonovich (red dashed line with error regions) approximations for the FBM with $H=0.6$ and $H=0.8$.
The black curve was computed by the Cholesky method. The  errors are shown because the integrals
were computed by a Monte Carlo method. A Kolmogorov--Smirnov statistics
for FPT distributions derived from Cholesky method in comparison to the Hertz approximation and Cholesky method in comparison to the Stratonovich approximation, yields the values  0.433 (p-value=1.03 $\times 10^{-6}$ ), 0.221 (p-value=0.655) and 0.294 (p-value=1.24 $\times 10^{-6}$), 0.256 (p-value=0.447) for H=0.6 and H=0.8, respectively.
}}
\label{fig:fig3}
\end{figure}
\end{center}

\newpage

\begin{center}
\begin{figure}
\includegraphics[width=0.6\columnwidth]{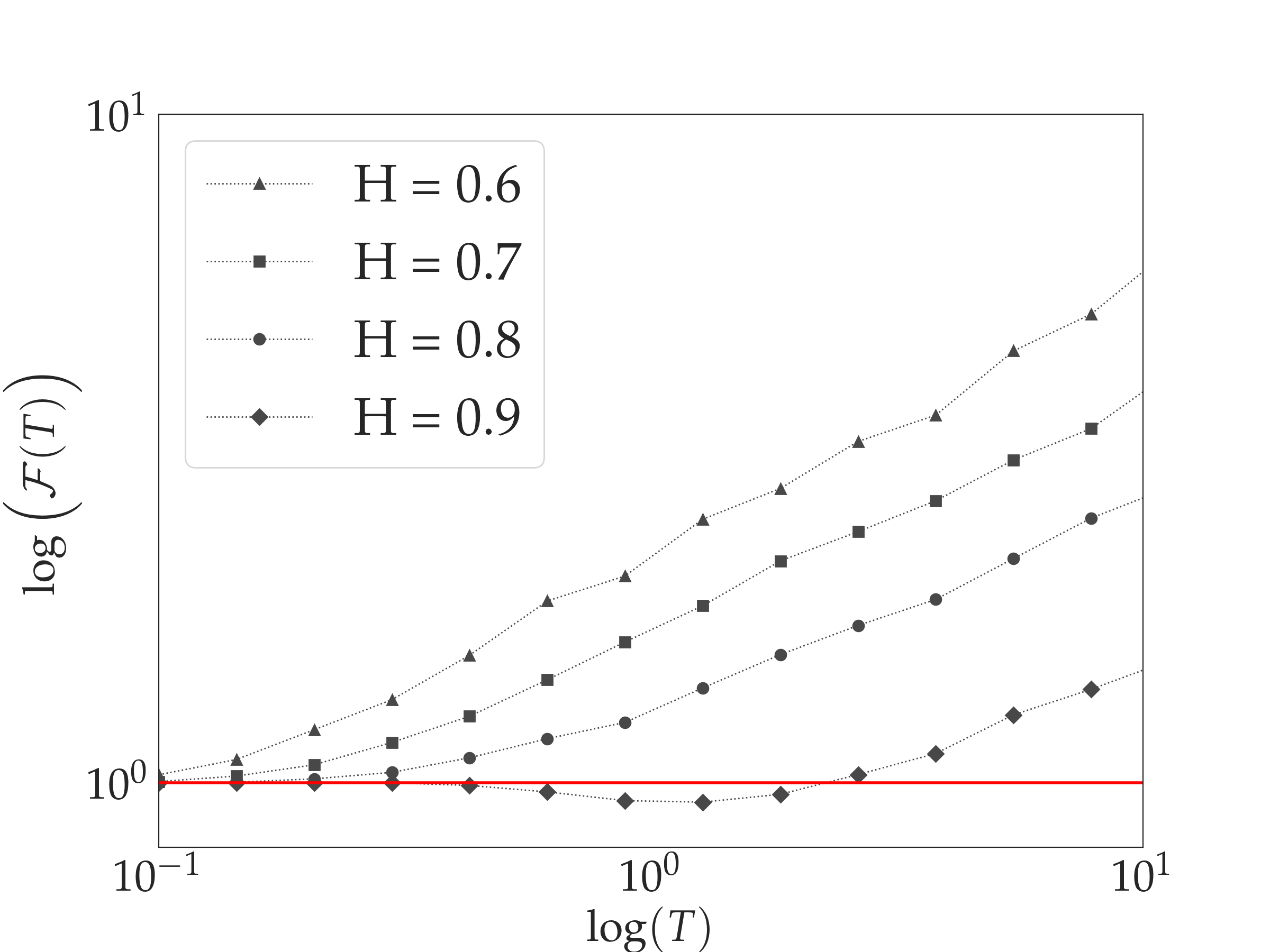}
\caption{\footnotesize{The Fano factor for up-crossings of the trajectories of the FBM with a Hurst
exponent $H$, as a function of the window size $T$. In the long-time limit, the up-crossing point
processes are slightly over-dispersed, whereas over short time scales, the Fano factor is equal to unity (red solid line),
a hallmark of the Poisson process, and the variance of the up-crossing over such short time scales
is equal to the mean.}}
\label{fig:fig4}
\end{figure}
\end{center}

\end{document}